\documentclass[preprint,pteplogo]{ptephy_v2}  % to generate preprint number with ptep logo
\preprintnumber{XXXX-XXXX}

\usepackage{subcaption}
\usepackage{amsmath}  % for dealing with mathematics,
\usepackage{amsthm}  % for dealing with theorem environments,
\usepackage{hyperref}  % for linking the cross references
\usepackage{lineno}

\usepackage{bm,epstopdf,natbib,hyperref,color,verbatim,multirow,bm,tikz,xcolor}
\hypersetup{colorlinks=true,urlcolor=blue,citecolor=blue,linkcolor=blue,menucolor=blue,anchorcolor=blue,filecolor=blue}
\everymath{\displaystyle}
\definecolor{lime}{HTML}{A6CE39}
\DeclareRobustCommand{\orcidicon}{
  \begin{tikzpicture}  
    \draw[lime, fill=lime] (0,0) 
    circle [radius=0.16] 
    node[white] {{\fontfamily{qag}\selectfont \tiny ID}};
    \draw[white, fill=white] (-0.0625,0.095) 
    circle [radius=0.007];
  \end{tikzpicture}
  \hspace{-3mm}
}
\foreach \x in {A, ..., Z}{\expandafter\xdef\csname orcid\x\endcsname{\noexpand\href{https://orcid.org/\csname orcidauthor\x\endcsname}
  {\noexpand\orcidicon}}
}

\begin{document}
%%%%%%%%%%%%%%%%%%%%%%%%%%%%%%%%%%%%%%%%%%%%%%%%%%%%%%%%%%%%%%%%%%%%%%%%%%%%%%%%

%%%%%%%%%%%%%%%%%%%%%%%%%%%%%%%%%%%%%%%%%%%%%%%%%%%%%%%%%%%%%%%%%%%%%%%%%%%%%%%%
\title{Afterpulse prediction for SUBMET experiment}
%%%%%%%%%%%%%%%%%%%%%%%%%%%%%%%%%%%%%%%%%%%%%%%%%%%%%%%%%%%%%%%%%%%%%%%%%%%%%%%%

%%%%%%%%%%%%%%%%%%%%%%%%%%%%%%%%%%%%%%%%%%%%%%%%%%%%%%%%%%%%%%%%%%%%%%%%%%%%%%%%
%   Authors
% To generate auto affiliation numbers please use \author{}\affil{} command
%%%%%%%%%%%%%%%%%%%%%%%%%%%%%%%%%%%%%%%%%%%%%%%%%%%%%%%%%%%%%%%%%%%%%%%%%%%%%%%%
\author[1]{Claudio Campagnari\hspace{-1mm}\orcidP{}}
\author[2]{Sungwoong Cho\footnote{Present address: Department of General Studies, Hongik University, 22639 Sejong-ro, Jochiwon-eup, Sejong, Korea}}
\author[2]{Suyong Choi\hspace{-1mm}\orcidG{}}
\author[2]{Seokju Chung\hspace{-1mm}\orcidH{}\footnote{Present address: Columbia University, 116th and Broadway, New York, New York 10027, USA}}
\author[3]{Matthew Citron\hspace{-1mm}\orcidF{}}
\author[6]{Ryan De Los Santos\hspace{-1mm}\orcidS{}}
\author[4]{Albert De Roeck\hspace{-1mm}\orcidI{}}
\author[4]{Martin Gastal}
\author[2]{Seungkyu Ha\hspace{-1mm}\orcidO{}}
\author[5]{Andy Haas}
\author[6]{Christopher Scott Hill\hspace{-1mm}\orcidJ{}}
\author[2]{Byeong Jin Hong}
\author[2]{Haeyun Hwang}
\author[2]{Insung Hwang\hspace{-1mm}\orcidD{}\footnote{Present address: Boston University, Commonwealth Ave, Boston, Massachusetts 02215, USA}}
\author[2]{Hoyong Jeong\hspace{-1mm}\orcidB{}}
\author[2]{Minseo Kim\hspace{-1mm}\orcidT{}}
\author[2]{Hyunki Moon\hspace{-1mm}\orcidC{}}
\author[2]{Jayashri Padmanaban\hspace{-1mm}\orcidK{}}
\author[1]{Ryan Schmitz\hspace{-1mm}\orcidE{}}
\author[2]{Changhyun Seo}
\author[1]{David Stuart\hspace{-1mm}\orcidL{}}
\author[3]{Juan Salvador Tafoya Vargas\hspace{-1mm}\orcidR{}}
\author[2]{Eunil Won\hspace{-1mm}\orcidM{}}
\author[2]{Jae Hyeok Yoo\hspace{-1mm}\orcidA{}\footnote{Email: jaehyeokyoo@korea.ac.kr}}
\author[2]{Jinseok Yoo\hspace{-1mm}\orcidQ{}\footnote{Email: lastbartican825@korea.ac.kr}}
\author[7]{Ayman Youssef}
\author[7]{Ahmad Zaraket}
\author[7]{Haitham Zaraket\hspace{-1mm}\orcidN{}}
\author[6]{Collin Zheng}

\affil[1]{Department of Physics, University of California, Santa Barbara, California 93106, USA}
\affil[2]{Department of Physics, Korea University, 145 Anam-ro, Seongbuk-gu, Seoul 02841, Korea}  
\affil[3]{Department of Physics, University of California, One Shields Avenue, Davis, California 95616, USA}
\affil[4]{CERN, CH-1211 Geneva, Switzerland}
\affil[5]{Department of Physics, New York University, 726 Broadway, New York, New York 10012, USA} 
\affil[6]{Department of Physics, The Ohio State University, 191 West Woodruff Ave, Columbus, Ohio 43210, USA}
\affil[7]{Multidisciplinary Physics Lab, Lebanese University, RGHC+4PR, Hadeth-Beirut, Lebanon}
%\email{xxxx@xxxx.ac.jp}}

%\author{Insert second author name here}
%\affil{Insert second author address here}

%\author{Insert third author name here}
%\author[3]{Insert fourth author name here} %%% Use optional bracket [3] to change the respective address
%\affil{Insert third author address here}

%\author{Insert last author name here\thanks{These authors contributed equally to this work}}
%\affil{Insert last author address here}

%%% To include the collaborator name... Please use the command "\collaborator"
%%% For example: \collaborator{ATLAS Collaboration}

%%%%%%%%%%%%%%%%%%%%%%%%%%%%%%%%%%%%%%%%%%%%%%%%%%%%%%%%%%%%%%%%%%%%%%%%%%%%%%%%
\begin{abstract}%
%%%%%%%%%%%%%%%%%%%%%%%%%%%%%%%%%%%%%%%%%%%%%%%%%%%%%%%%%%%%%%%%%%%%%%%%%%%%%%%%
The SUB-Millicharge ExperimenT (SUBMET) investigates an unexplored parameter space of millicharged particles with mass $m_\chi < $ 1.6 GeV/c$^2$ and charge $Q_\chi < 10^{-3}e$. The detector consists of an Eljen-200 plastic scintillator coupled to a Hamamatsu Photonics R7725 photomultiplier tube (PMT). PMT afterpulses, delayed pulses produced after an energetic pulse, have been observed in the SUBMET readout system, especially following primary pulses with a large area. We present a prediction method for afterpulse rates based on measurable parameters, which reproduces the observed rate with approximately 20\% precision. This approach enables a better understanding of afterpulse contributions and, consequently, improves the reliability of background predictions.
%%%%%%%%%%%%%%%%%%%%%%%%%%%%%%%%%%%%%%%%%%%%%%%%%%%%%%%%%%%%%%%%%%%%%%%%%%%%%%%%
\end{abstract}
%%%%%%%%%%%%%%%%%%%%%%%%%%%%%%%%%%%%%%%%%%%%%%%%%%%%%%%%%%%%%%%%%%%%%%%%%%%%%%%%

%-------------------------------------------------------------------------------
\subjectindex{xxxx, xxx}
\maketitle
%-------------------------------------------------------------------------------

%\linenumbers
%%%%%%%%%%%%%%%%%%%%%%%%%%%%%%%%%%%%%%%%%%%%%%%%%%%%%%%%%%%%%%%%%%%%%%%%%%%%%%%%
\section{Introduction}\label{sec:intro}
%%%%%%%%%%%%%%%%%%%%%%%%%%%%%%%%%%%%%%%%%%%%%%%%%%%%%%%%%%%%%%%%%%%%%%%%%%%%%%%%

%%%%%%%%%%%%%%%%%%%%%%%%%%%%%%%%%%%%%%%%%%%%%%%%%%%%%%%%%%%%
%\subsection{Overview of SUBMET}\label{sec:intro:overview}
%%%%%%%%%%%%%%%%%%%%%%%%%%%%%%%%%%%%%%%%%%%%%%%%%%%%%%%%%%%%
The existence of dark matter has been supported by a variety of astrophysical and cosmological observations, yet its direct detection remains an experimental challenge. One of the dark matter candidates is the millicharged particle ($\chi$), which arises in scenarios that introduce a new $U(1)$ in the dark sector with a massless dark-photon and a massive dark-fermion ($\chi$)~\cite{HOLDOM1986196,PhysRevD.79.015014,POSPELOV2009391}.

\begin{figure}[h!]
%---------------------------------------
\centering
    \includegraphics[width=.5\linewidth]{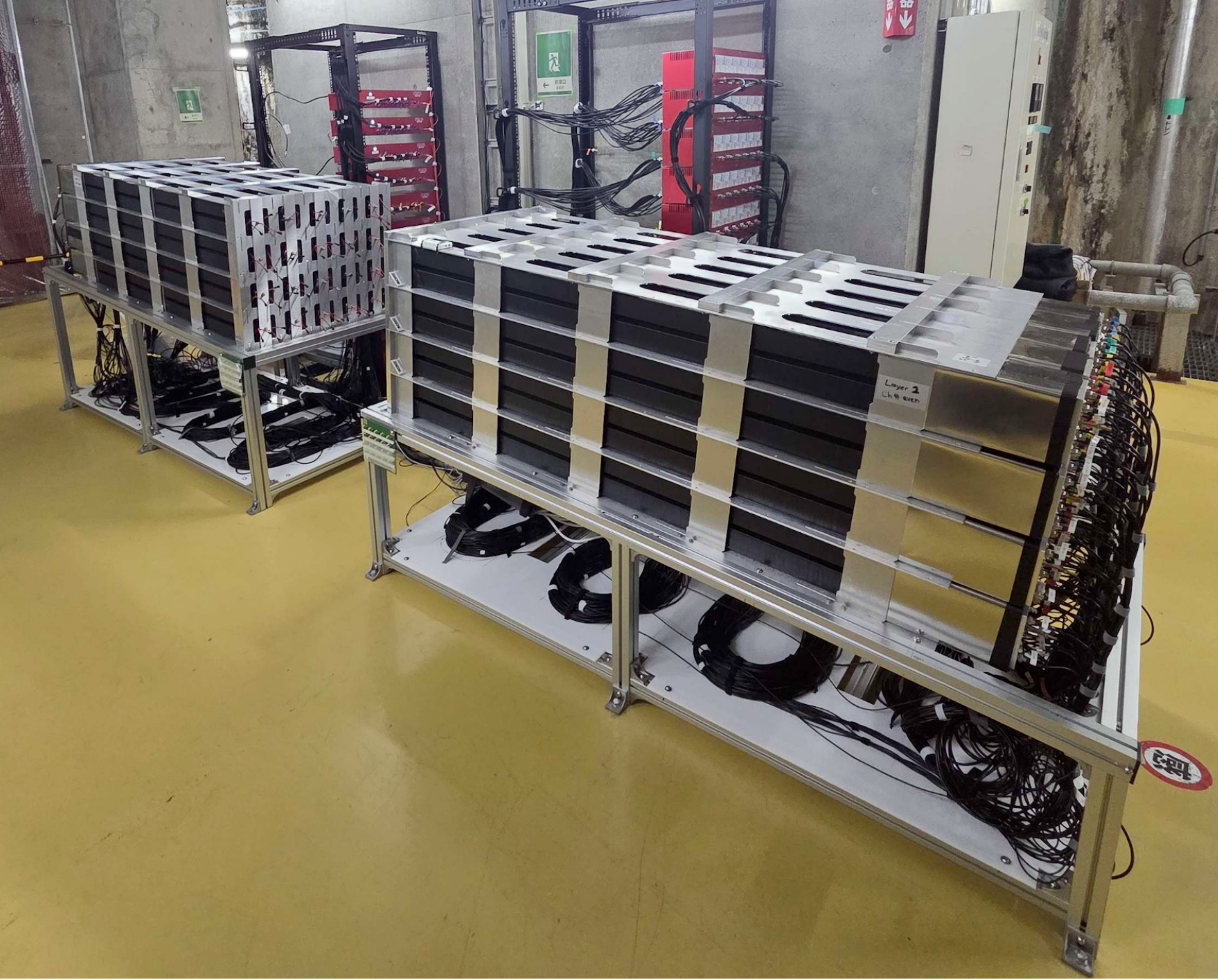}
    \caption{The SUBMET detector installed at the J-PARC neutrino monitor building.}
    \label{fig:intro:detector}
%---------------------------------------
\end{figure}

The SUB-Millicharge ExperimenT (SUBMET)~\cite{Kim:2021eix} aims to search for millicharged particles using 30 GeV proton-target collisions at J-PARC (Japan Proton Accelerator Research Complex). The detector is designed to probe the previously unexplored parameter space of mass $m_\chi < 1.6 \text{~GeV/c}^2$ and charge $Q_\chi < 10^{-3}e$. It is currently installed at J-PARC, as shown in Figure~\ref{fig:intro:detector}, and has been collecting data since 2024. The detector is composed of 160 ``modules"~\cite{mechanics}, each consisting of an Eljen-200 plastic scintillator bar~\cite{ej200} coupled with a Hamamatsu Photonics R7725 photomultiplier tube (PMT)~\cite{r7725}. 
%---------------------------------------
%---------------------------------------
%%%%%%%%%%%%%%%%%%%%%%%%%%%%%%%%%%%%%%%%
%\subsection{Motivation of this study}\label{sec:intro:motiv}
%%%%%%%%%%%%%%%%%%%%%%%%%%%%%%%%%%%%%%%%

When photoelectrons produced in the PMT ionize gas impurities between the photocathode and dynodes, these ionized gases can also generate photoelectrons at the photocathode a few hundred nanoseconds later. These delayed signals are known as ``afterpulses."% and it has been studied in other researches~\cite{ionap, KM3NeT:2018nwm, afterpulse}.
The afterpulse effect has also been observed in the SUBMET detector, especially within 720--3600~ns after energetic particles pass through the PMT. The detector response to energetic particles is observed as a ``large pulse" (pulses with a height greater than 140~mV) in our readout system.
In the SUBMET detector, the signatures of afterpulses and our target signal, single photoelectron (SPE) pulses, are typically indistinguishable, which can increase the background rate.
%Because both afterpulses and our signal---single photoelectron (SPE) pulses---both originate from a single electron, the signatures of these two pulses are indistinguishable, which can increase the background rate. 
The 30 GeV J-PARC proton beam consists of eight bunches separated by $\sim 600$ ns~\cite{Friend_2017}. When a large pulse is generated in a bunch, a number of afterpulses are produced and may overlap with SPE signals in the subsequent bunches.
%In the time of interest, when 8 proton bunches collide with the graphite target during the J-PARC fast extraction (FX) beam operation (581~ns interval, 80~ns width, within a 1.36~s cycle~\cite{beam}), afterpulses may overlap with signals from subsequent bunches.

To study the effect of afterpulses in the SUBMET detector, we collected beam data in June 2024 using a dedicated trigger delay, focusing on the time window following the last bunch. Based on these data, we developed a model to predict the afterpulse rate at a given time. This prediction method enables us to include events that contain large pulses in our data analysis by precisely estimating the contributions from afterpulses.   

This paper is organized as follows. Section~\ref{sec:method} outlines the steps required to establish the prediction model. The key observables, the afterpulse counts as a function of the large pulse area and the time distribution of the afterpulses, are described in Sections~\ref{sec:method:desc:depend} and~\ref{sec:method:desc:timedist}. Section ~\ref{sec:method:desc:model} introduces a prediction formula based on these observables. The measurement of prediction parameters is described in Section~\ref{sec:method:params}, and performance of the prediction model is presented in Section~\ref{sec:method:perf}.

%%%%%%%%%%%%%%%%%%%%%%%%%%%%%%%%%%%%%%%%%%%%%%%%%%%%%%%%%%%%%%%%%%%%%%%%%%%%%%%%
\section{Method}\label{sec:method}
\begin{figure}
%---------------------------------------
\centering
\begin{subfigure}[b]{.49\linewidth}
    \centering
    \includegraphics[width=\linewidth]{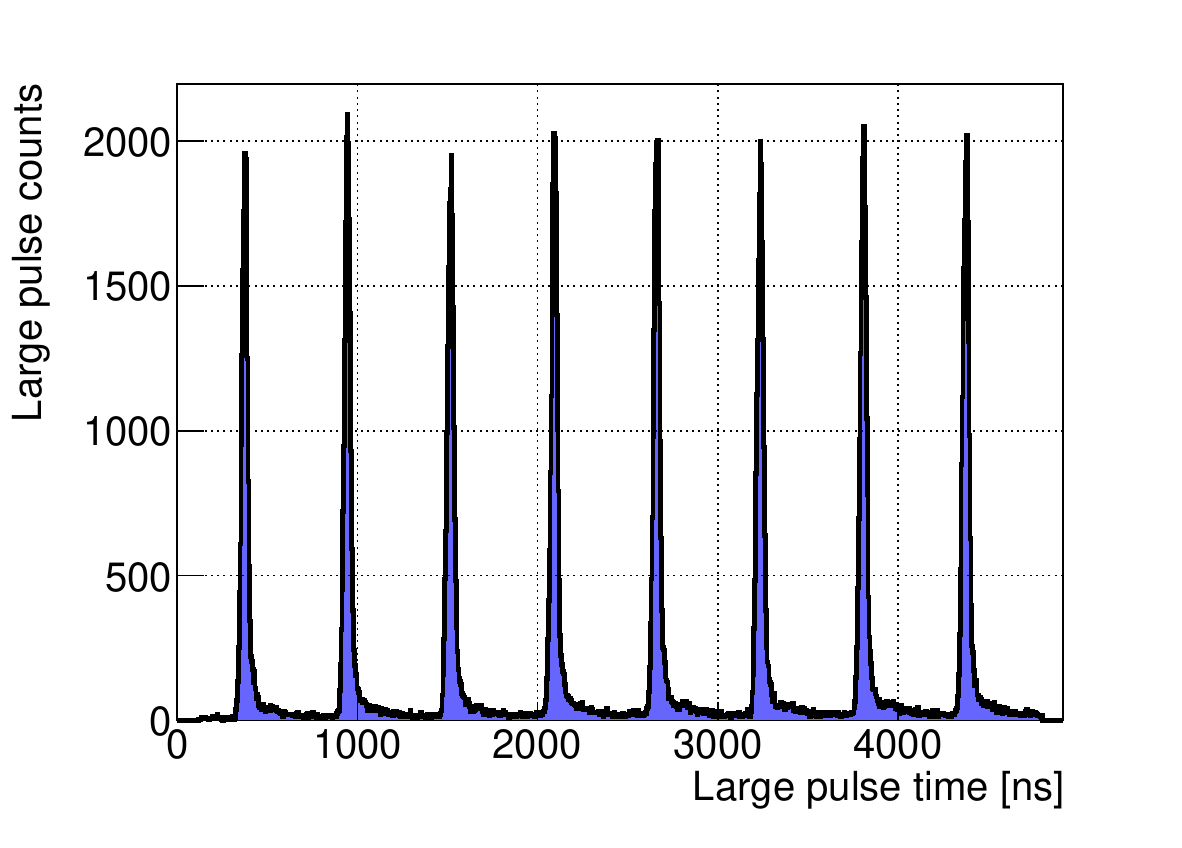}
    \caption{}
    \label{fig:beamstructure:fullbunch}
\end{subfigure}
\hfill
\begin{subfigure}[b]{.49\linewidth}
    \centering
    \includegraphics[width=\linewidth]{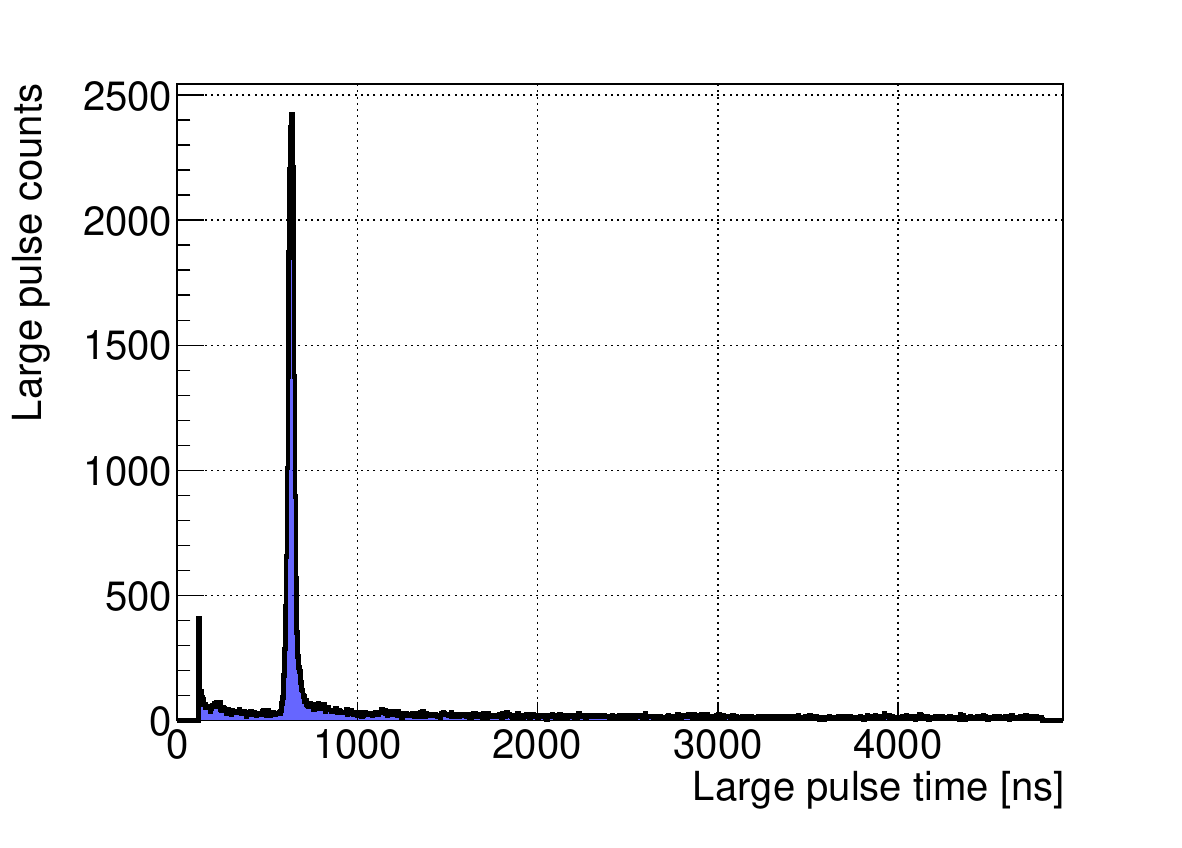}
    \caption{}
    \label{fig:beamstructure:aprun}
\end{subfigure}
\caption{Time distribution of pulses with heights exceeding 140~mV. (a) shows the full beam structure consisting of 8 bunches as recorded in the standard data-taking mode. (b) shows a delayed time window in the run dedicated to afterpulse studies.}
\label{fig:beamstructure}
%---------------------------------------
\end{figure}

%%%%%%%%%%%%%%%%%%%%%%%%%%%%%%%%%%%%%%%%%%%%%%%%%%%%%%%%%%%%%%%%%%%%%%%%%%%%%%%%
The duration of the SUBMET readout window is 5~\textmu s. Normally, the trigger delay is set so that all eight bunches are contained the window as shown in Figure~\ref{fig:beamstructure:fullbunch}. However, to study the afterpulse effect, it was adjusted so that the readout window includes the 4~\textmu s after the last bunch. The beam structure recorded by the adjusted time setting is shown in Figure~\ref{fig:beamstructure:aprun}. We utilized 50k events, with one half used to fit the prediction model parameters and the remaining half to assess its performance. 
%---------------------------------------
%---------------------------------------

\begin{figure}[h!]
%---------------------------------------
\centering
\includegraphics[width=\linewidth]{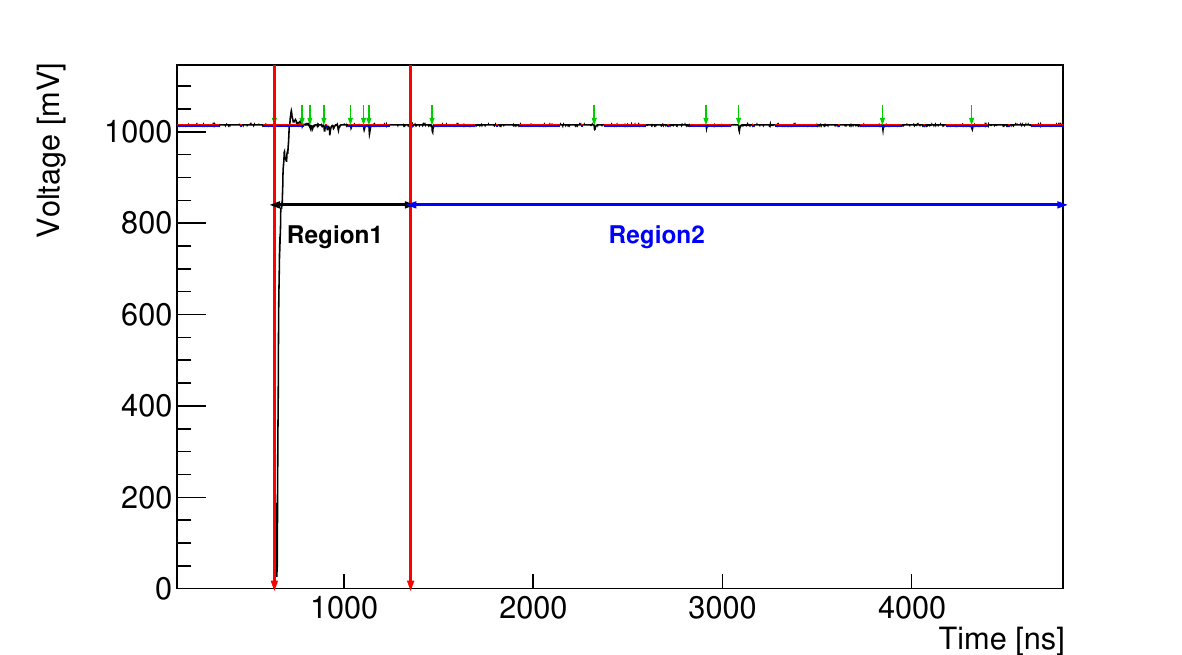}
\caption{Waveform of an afterpulse event. Green arrows indicate the detected pulses by the SUBMET pulse finding algorithm. A significant baseline fluctuation is observed immediately after the large pulse (Region1).}
\label{fig:APevt}
%---------------------------------------
\end{figure}

Figure~\ref{fig:APevt} shows a waveform from a channel in which a large pulse is observed at 600 ns, followed by smaller pulses, predominantly single photoelectrons (SPEs). A significant fraction of the pulses in Region1 (defined as the 720 ns window following the large pulse) are attributed to the fluctuation of the PMT signal. Consequently, this region is excluded from the development of the prediction model.

%---------------------------------------
%---------------------------------------

%%%%%%%%%%%%%%%%%%%%%%%%%%%%%%%%%%%%%%%%%%%%%%%%%%%%%%%%%%%%
\subsection{General description about the method}\label{sec:method:desc}
%%%%%%%%%%%%%%%%%%%%%%%%%%%%%%%%%%%%%%%%%%%%%%%%%%%%%%%%%%%%
The dark current and radiation rates were measured using data collected during beam-off periods. The dark current and radiation rates are in $\mathcal{O}(10^3)$~Hz, whereas the afterpulse rate is in $\mathcal{O}(10^6)$~Hz. Therefore, contributions from dark current and radiation are negligible when measuring the afterpulse rates. It should also be noted that afterpulses produced by earlier bunches can contribute to the background. To ensure that the afterpulses within the selected time window predominantly arise from the last bunch, events containing any pulses prior to the last bunch timing are rejected.

The prediction model is constructed under the assumption that the number of afterpulses per unit time ($dn/dt$) or afterpulse rate depends on the area ($A$) of the large pulse that generates afterpulses and the time constant of afterpulse rate ($\tau$). Since $\tau$ varies by modules, it is obtained  individually for each of them. 

%%%%%%%%%%%%%%%%%%%%%%%%%%%%%%%%%%%%%%%%
\subsubsection{The dependence of afterpulse counts on the large pulse area}\label{sec:method:desc:depend}
\begin{figure}[]
%---------------------------------------
\centering
    \centering
     \includegraphics[width=.8\linewidth]{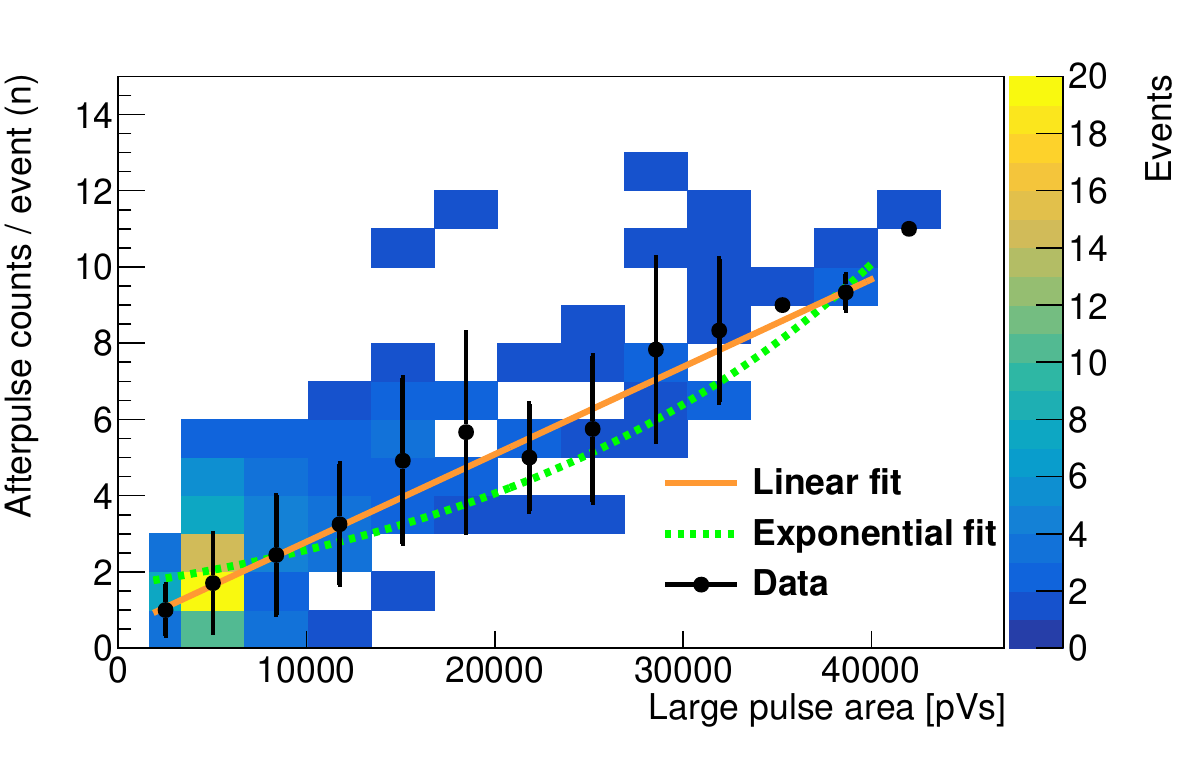}
    \caption{Distribution of number of afterpulses $n$ vs pulase area $A$. The black dots represent the average number of afterpulses in each area bin. The orange solid and green dashed lines correspond to the linear and the exponential fits, respectively.}
    \label{fig:dependency:linandexpo}
\label{fig:dependency}
%---------------------------------------
\end{figure}

%%%%%%%%%%%%%%%%%%%%%%%%%%%%%%%%%%%%%%%%
The number of afterpulses in a single event~($n$) is measured with varying~$A$ for each module within a fixed time window of interest ($t \in [t_0 = 720~\text{ns}, t_1 = 3475.2~\text{ns}]$). Figure~\ref{fig:dependency} illustrates a positive correlation between $A$ and $n$. This distribution is fitted using linear and exponential functions of $A$ as follows.
%---------------------------------------
\begin{align}
%---------------------------------------
\text{linear fit: }\quad n_{\mathrm{pred}}^{\mathrm{lin}}(A| t_0, t_1)  &=   p_0^{\mathrm{lin}} + p_1^{\mathrm{lin}} A \label{eq:lin} \\
\text{exponential fit: } n_{\mathrm{pred}}^{\mathrm{expo}}(A| t_0, t_1) &= \exp(p_0^{\mathrm{expo}} + p_1^{\mathrm{expo}} A) \label{eq:expo}
%---------------------------------------
\end{align}
%---------------------------------------
where $p_0^{\mathrm{lin}}, p_1^{\mathrm{lin}}, p_0^{\mathrm{expo}}, \mathrm{p_1^{expo}}$ are fit parameters and $n_{\mathrm{pred}}^{\mathrm{lin}}(A| t_0, t_1), n_{\mathrm{pred}}^{\mathrm{expo}}(A| t_0, t_1)$ are the expected number of afterpulses between $t_0$ and $t_1$ for a given $A$, based on linear and exponential fits, respectively. %Among Eq.~\ref{eq:lin} and Eq.~\ref{eq:expo}, the one that provides better agreement with the data is selected.
%---------------------------------------
%---------------------------------------

%%%%%%%%%%%%%%%%%%%%%%%%%%%%%%%%%%%%%%%%
\subsubsection{Afterpulse Time Structure}\label{sec:method:desc:timedist}
%%%%%%%%%%%%%%%%%%%%%%%%%%%%%%%%%%%%%%%%
%The afterpulse count in Section~\ref{sec:method:desc:depend} is measured within a fixed time window. To apply our prediction to the time window of interest, the time structure of afterpulses must be taken into account. 

\begin{figure}[]
%---------------------------------------
\centering 
\includegraphics[width=.8\linewidth]{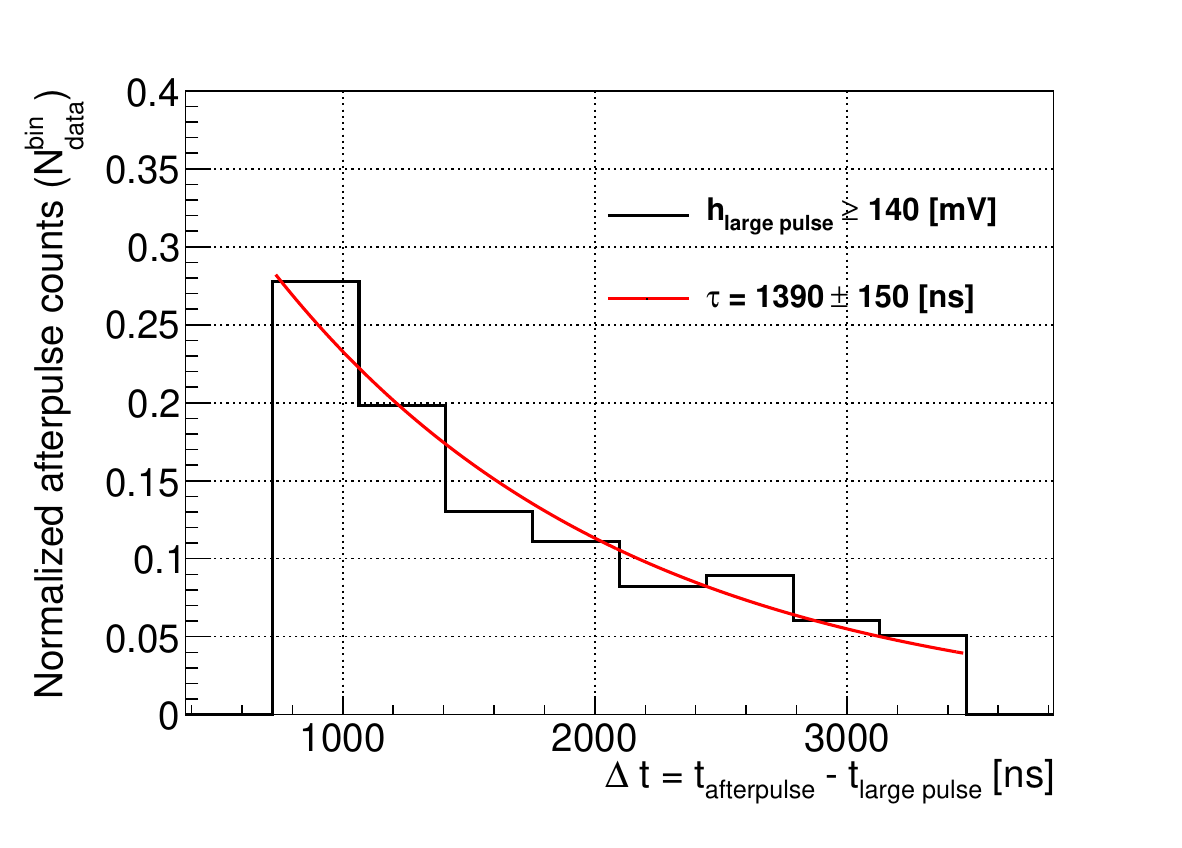}
\caption{Distribution of $\Delta t = t_\textrm{afterpulse}-t_\textrm{large pulse}$ normalized to unit area.}
\label{fig:time}
%---------------------------------------
\end{figure}

As shown in Figure~\ref{fig:time}, the number of afterpulses in each time bin ($N^{\mathrm{bin}}_{\mathrm{data}}$) decreases exponentially during the time of interest as the time difference between the large pulse and an afterpulse ($\Delta t = t_\textrm{afterpulse}-t_\textrm{large pulse}$) increases. The time constant~($\tau$) of the afterpulse rate for each module is determined by fitting the time difference distribution to an exponential function. Dependence of $\tau$ on $A$ was studied and found to be negligible. Therefore, $\tau$ was measured independently of $A$.

%---------------------------------------
%---------------------------------------

%%%%%%%%%%%%%%%%%%%%%%%%%%%%%%%%%%%%%%%%
\subsubsection{Prediction model}\label{sec:method:desc:model}
%%%%%%%%%%%%%%%%%%%%%%%%%%%%%%%%%%%%%%%%
Based on the dependence of afterpulse rates on the area of the large pulse and the structure of afterpulse time, the prediction model is established as follows. Due to the exponentially decreasing number of afterpulses within the time of interest and the negligible correlation between $\tau$ and $A$, $dn/dt$ can be expressed as
%Due to the exponentially decaying time structure of afterpulses and the negligible correlation between $\tau$ and $A$, $dn/dt$ can be expressed as
%---------------------------------------
\begin{align}
%---------------------------------------
    \dfrac{dn}{dt} = \dfrac{1}{\tau}n_0(A)\exp(-t/\tau)\label{eq:time},
%---------------------------------------
\end{align}
%---------------------------------------
where
%---------------------------------------
\begin{align}
%---------------------------------------
     n_0(A) = \dfrac{n_{\mathrm{pred}}^{\mathrm{lin}}(A| t_0, t_1)}{\exp(-t_0 / \tau ) - \exp(-t_1 / \tau)} \qquad \text{or} \qquad \dfrac{n_{\mathrm{pred}}^{\mathrm{expo}}(A| t_0, t_1)}{\exp(-t_0 / \tau ) - \exp(-t_1 / \tau)}
%---------------------------------------
\end{align}
%---------------------------------------
is determined by integrating Eq.~\ref{eq:time} from $t_0$ to $t_1$ and equating it with Eqs.~\eqref{eq:lin} and~\eqref{eq:expo}.

Then, the predicted number of afterpulses within an arbitrary time window $t \in [t_i, t_f]$ with $t_i \geq 720 \text{~ns}$, denoted as $n_{\mathrm{pred}}(A, t_i, t_f)$, is calculated using the measured parameters $p_0, p_1$ and $\tau$ as 
%---------------------------------------
\begin{align}
%---------------------------------------
n_\mathrm{pred}(A, t_i, t_f) = \dfrac{1}{\tau} \int_{t_i}^{t_f}n_0(A) \exp(-t' / \tau) dt' = 
\begin{cases}
    (p_0^{\mathrm{lin}} + p_1^{\mathrm{lin}} A) \times \dfrac{e^{- t_i / \tau} - e^{-t_f/ \tau}}{e^{-t_0 / \tau} - e^{-t_1 / \tau}} \\
    \exp{(p_0^{\mathrm{expo}} + p_1^{\mathrm{expo}} A)}\times \dfrac{e^{- t_i / \tau} - e^{-t_f/ \tau}}{e^{-t_0 / \tau} - e^{-t_1 / \tau}}.
\end{cases}
\label{eq:prediction}
%---------------------------------------
\end{align}
%---------------------------------------
 Hereafter, the prediction model using $n_{\mathrm{pred}}^{\mathrm{lin}}(A| t_0, t_1) = p_0^{\mathrm{lin}} + p_1^{\mathrm{lin}} A$ is referred to as the linear prediction model, while that using $n_{\mathrm{pred}}^{\mathrm{expo}}(A| t_0, t_1) = \exp{(p_0^{\mathrm{expo}} + p_1^{\mathrm{expo}} A)}$ is referred to as the exponential prediction model.

%%%%%%%%%%%%%%%%%%%%%%%%%%%%%%%%%%%%%%%%%%%%%%%%%%%%%%%%%%%%
\subsection{Measurement of parameters}\label{sec:method:params}
%%%%%%%%%%%%%%%%%%%%%%%%%%%%%%%%%%%%%%%%%%%%%%%%%%%%%%%%%%%%

The value of parameters, $p_0, p_1$ and $\tau$, required for the prediction differs for each module, so we performed independent measurements of them. The parameters are optimized for each module, while the prediction model itself remains unchanged. %This section concludes with summary of the measured parameters.
%SPE count vs large pulse area

As described in Section~\ref{sec:method:desc:depend}, the parameters $p_0$, $p_1$ were obtained by fitting linear or exponential functions to the distribution of afterpulse counts versus large pulse area. The results of the $p_0$ and $p_1$ measurements are summarized in Table~\ref{parameter}.
\begin{table}[!h]
\caption{A Summary of $p_0$ and $p_1$ for linear and exponential prediction model.}%%%Table caption goes here
\label{parameter}
\centering
\begin{tabular}{|c||c|c|}%%%The number of columns has to be defined here
\hline
      & linear model & exponential model\\
      \hline
$p_0$ & $<$ 4 & $<$ 2 \\ \hline
$p_1$ & 3--14 $\times 10^{-5} $& 1--4 $\times 10^{-5}$ \\ \hline
%$p_0$ & 0.14 $\pm$ 0.671 & 0.062 $\pm$ 0.3516  \\ %%%% Table body
%$p_1$ & $0.63 \times 10^{-4} \pm 0.129 \times 10^{-4}$ & $0.19\times %10^{-4} \pm 0.043 \times 10^{-4}$ \\%%%% Table body
\end{tabular}
\end{table}%%%End of the table
%Plot 1: SPE count vs time 
 
\begin{figure}[h!]
%---------------------------------------
\centering
    \includegraphics[width=.8\linewidth]{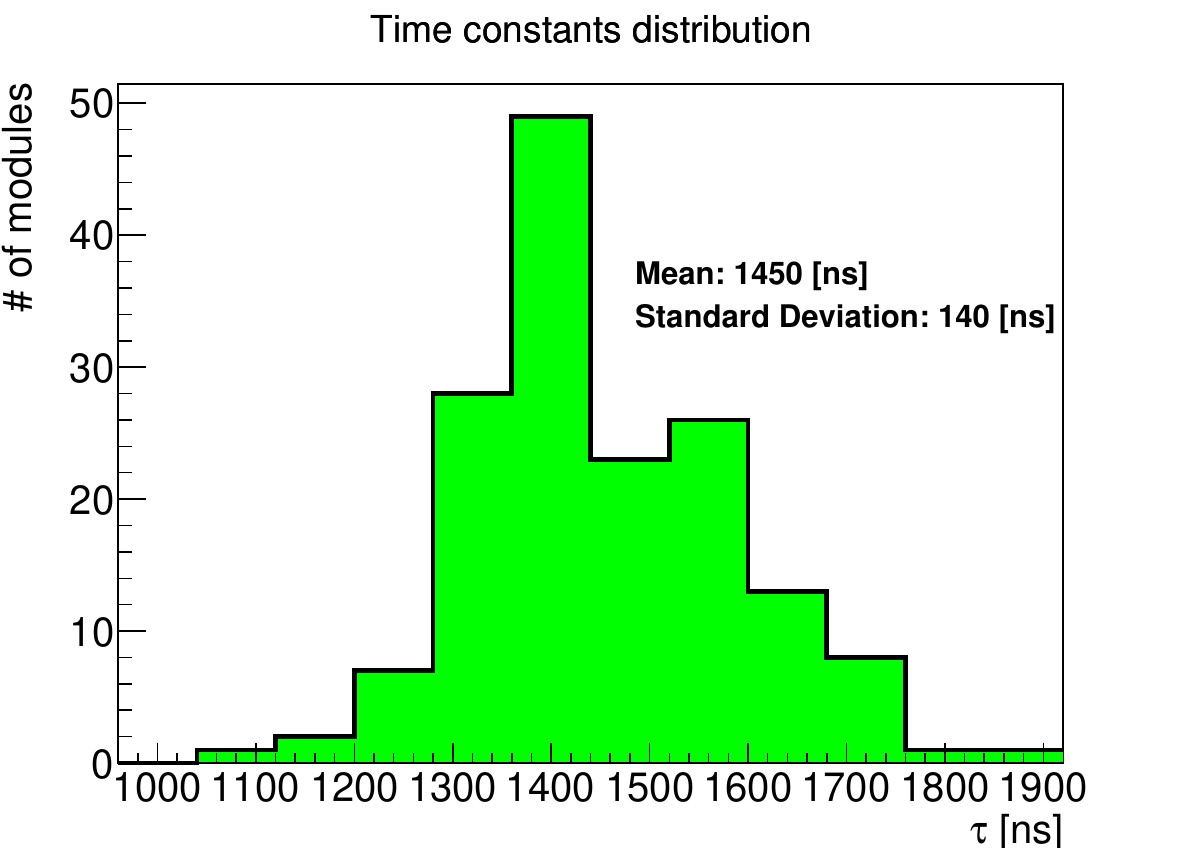}
    \caption{The distribution of $\tau$ for all modules.}
    \label{fig:tau}
%---------------------------------------
\end{figure}

The afterpulse time constants, $\tau$, are determined by fitting the distribution of time differences between large pulses and afterpulses to  $(\text{constant}) \times\exp(- t / \tau)$ as explained in~\ref{sec:method:desc:timedist}. The distribution of time constant $\tau$ for all modules is shown in Figure~\ref{fig:tau}.

%---------------------------------------
%---------------------------------------

%%%%%%%%%%%%%%%%%%%%%%%%%%%%%%%%%%%%%%%%%%%%%%%%%%%%%%%%%%%%
\subsection{Performance}\label{sec:method:perf}
\begin{figure}[h!]
%---------------------------------------
\centering
\includegraphics[width=\linewidth]{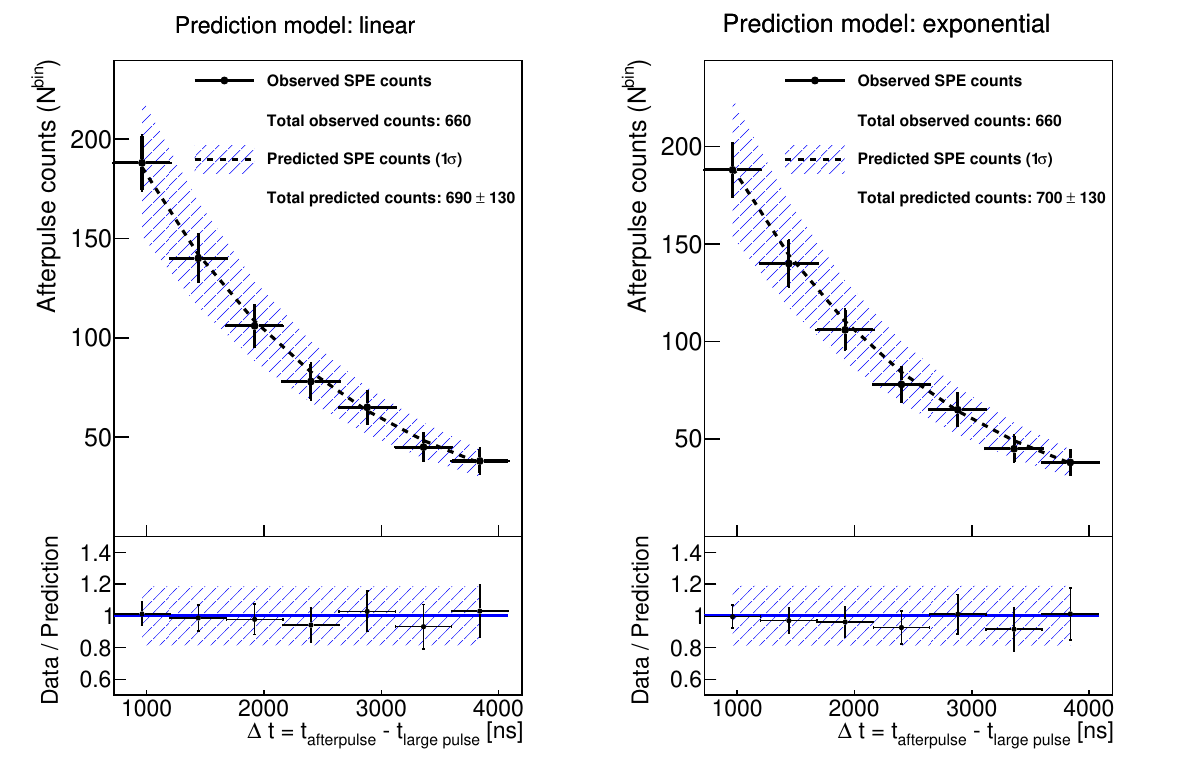}
\caption{Afterpulse (AP) time prediction with the linear (left) and exponential (right) models. Black circles represent observed counts, and blue dashed bands indicate the $\pm1\sigma$ band of predictions.}
\label{fig:pred_single}
%---------------------------------------
\end{figure}

%%%%%%%%%%%%%%%%%%%%%%%%%%%%%%%%%%%%%%%%%%%%%%%%%%%%%%%%%%%%
% Prediction vs data
The predicted number of afterpulses in each time bin ($N_{\mathrm{pred}}^{\mathrm{bin}}$) is obtained by integrating Eq.~\eqref{eq:time} over the bin range and accumulating the number of predicted afterpulse counts over all events. Figure~\ref{fig:pred_single} shows the SPE counts for a single channel in the test dataset as a function of $\Delta t$, along with predictions from the linear and the exponential fit models. The blue hatched area represents the 1$\sigma$ uncertainty band determined by the uncertainties on the parameters from the fit.

%---------------------------------------
\begin{figure}[]
%---------------------------------------
    \centering
    \includegraphics[width=1.0\linewidth]{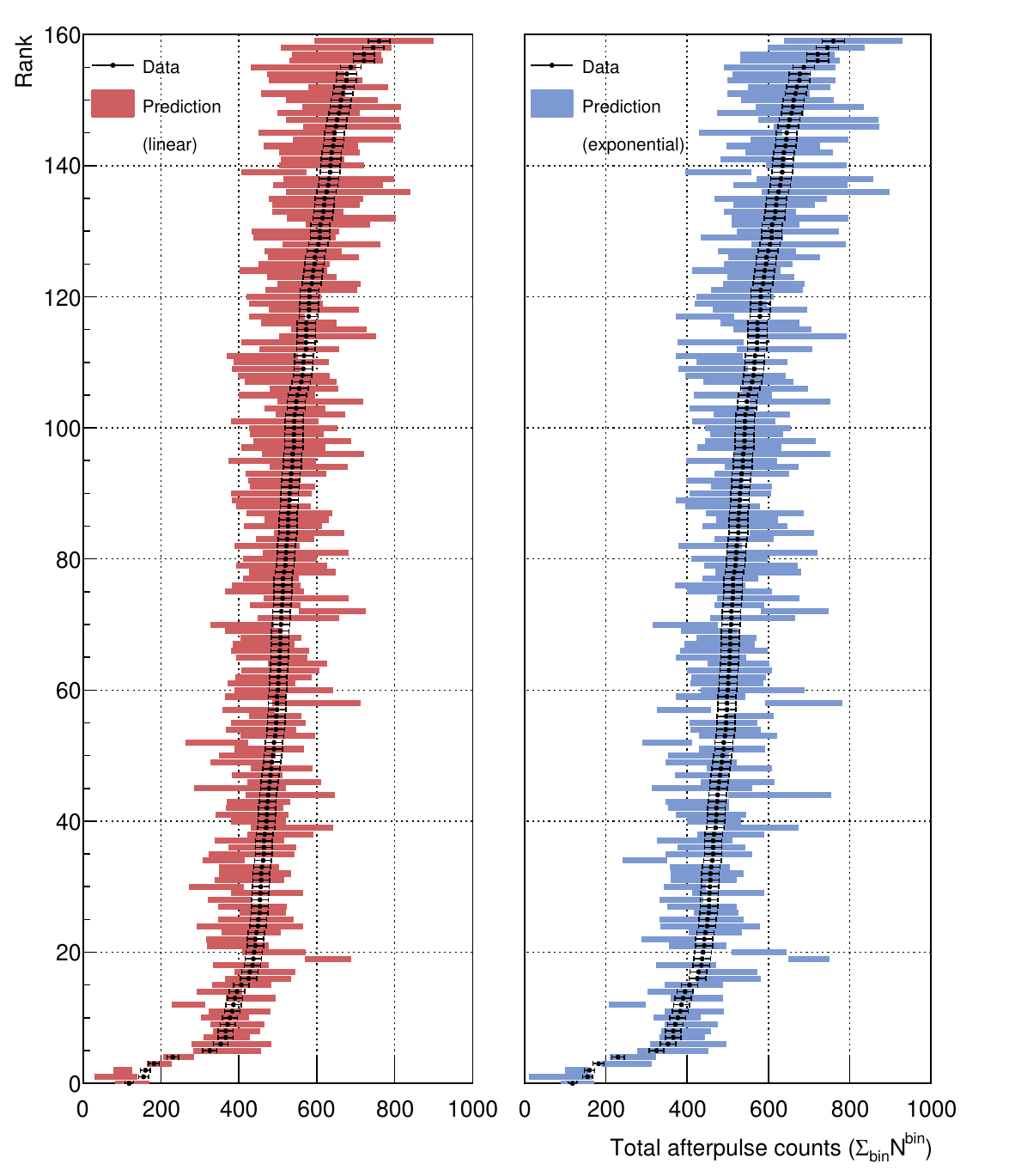}
    \caption{The results of prediction for 160 modules. Black dots represent the total observed afterpulse counts, and the error bars are the statistical uncertainties. Red and blue shaded regions represent predictions from the linear and exponential models, respectively, with their $1\sigma$ uncertainty bands.}
    \label{fig:pred_total}
%---------------------------------------
\end{figure}

%--------------------------------------- 
The afterpulse counts summed in the entire time bin and their prediction results $\displaystyle \left(\sum_{\text{bin}}N_{\text{pred}}^{\mathrm{bin}}\right)$ are summarized in Figure~\ref{fig:pred_total}, sorted in ascending order of the afterpulse counts. 5 Modules with relatively low afterpulse counts were found to have poor cable contact during the measurement.
%---------------------------------------
%---------------------------------------
%$\chi^2 / \text{ndf}$ distribution

\begin{figure}[!h]
%---------------------------------------
    \centering
    \includegraphics[width=\linewidth]{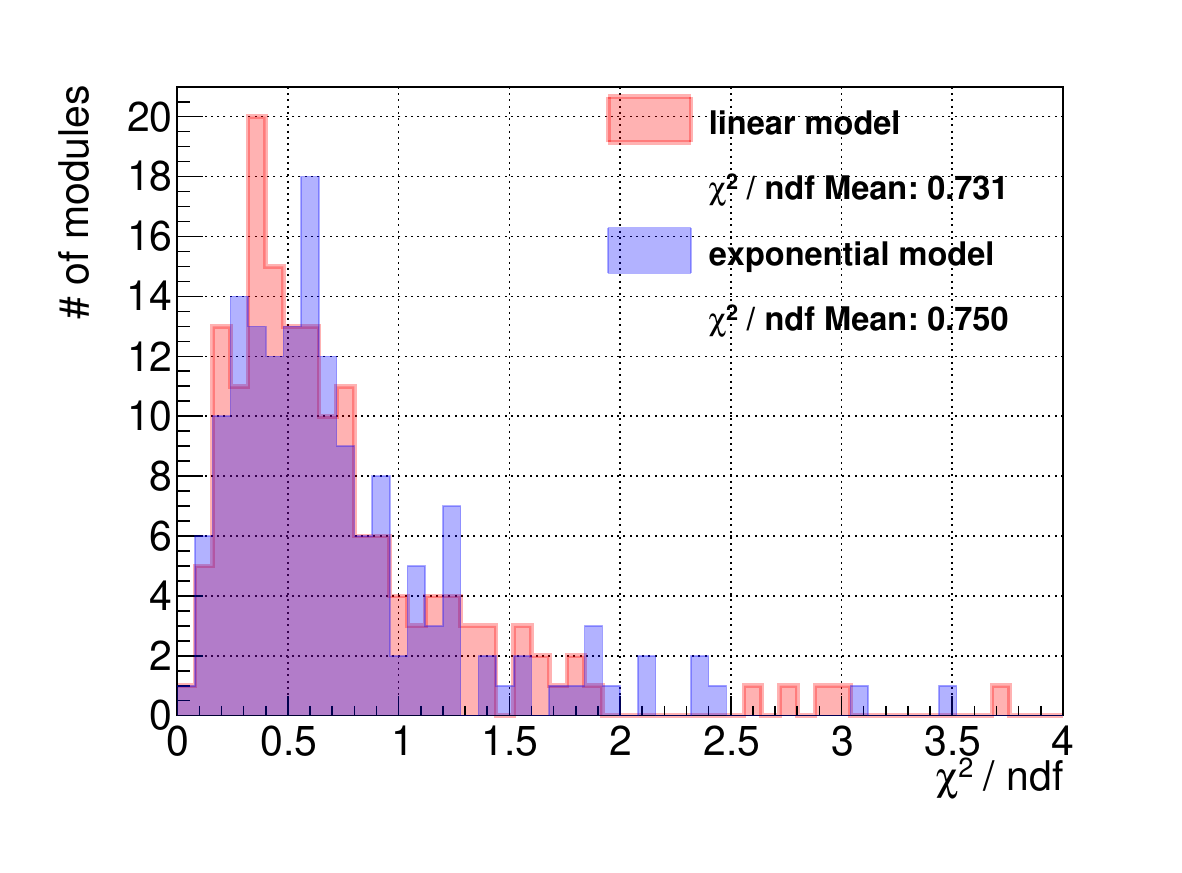}
    \caption{Distribution of $\chi^2 /\text{ndf}$ obtained from 160 modules. The red and blue distributions represent the linear and exponential prediction models, respectively.}
    \label{fig:chi}
%---------------------------------------    
\end{figure}

To evaluate the degree of agreement between prediction and measurements, we define the reduced $\chi^2$ as follows.
%---------------------------------------
\begin{align}
%---------------------------------------
    \chi^2 = \displaystyle \sum_{\mathrm{bin}} \dfrac{(N_\mathrm{data}^{\mathrm{bin}} - N_\mathrm{pred}^{\mathrm{bin}})^2}{(\sigma_\mathrm{data}^{\mathrm{bin}})^2 + (\sigma_{\mathrm{pred}}^{\mathrm{bin }})^2}
%---------------------------------------
\end{align}
%---------------------------------------
where $N_{\mathrm{data}}^{\mathrm{bin}}$ is the number of observed afterpulses within each time bin and $\sigma_{\mathrm{pred}}^{\mathrm{bin }}$ is the uncertainty on the prediction propagated from the uncertainties of the fit parameters. The typical value of $\sigma_{\mathrm{pred}}^{\mathrm{bin }}$ is about $20$\%. For linear and exponential prediction methods, $\chi^2 / \text{ndf}$ was calculated module by module, and the mean of the resulting distribution was obtained as shown in Figure~\ref{fig:chi}. Both models produced a similar mean $\chi^2/$ndf. For simplicity, the linear model is selected for further analysis, such as beam-on or beam-off background prediction. In addition, dependence on the height of a large pulse was also tested, but pulse area was found to be a more reliable variable for predicting afterpulse rates.

%---------------------------------------
%---------------------------------------

%%%%%%%%%%%%%%%%%%%%%%%%%%%%%%%%%%%%%%%%%%%%%%%%%%%%%%%%%%%%%%%%%%%%%%%%%%%%%%%%
\section{Summary and discussion}
%%%%%%%%%%%%%%%%%%%%%%%%%%%%%%%%%%%%%%%%%%%%%%%%%%%%%%%%%%%%%%%%%%%%%%%%%%%%%%%%

The SUB-Millicharge ExperimenT (SUBMET) at J-PARC offers a unique opportunity to search for millicharged particles in the region of parameter space previously unexplored. The detector consists of long plastic scintillator bars coupled to photomultiplier tubes (PMTs). Large PMT pulses are often followed by afterpulses that are nearly indistinguishable from single photoelectron (SPE) signals. Consequently, an accurate prediction of the afterpulse rate is crucial for reliable background estimation in the experiment.

We developed a method to predict the afterpulse rate using the area of a large pulse, responsible for generating afterpulses, and the timing of these afterpulses. The method exploits the correlation between the afterpulse count and the large-pulse area, as well as the exponentially decaying time distribution of afterpulse counts within the time window of interest. For each module, the afterpulse time constant and the correlation between the large-pulse area and the afterpulse count within a fixed time window were measured. Two predictive models were examined: one assuming a linear dependence of the afterpulse count on the large-pulse area, and the other assuming an exponential dependence. It is demonstrated that both models reproduce the observed data with comparable accuracy.

As more than 70\% of the collision events include at least one large pulse, characterizing afterpulse effects is crucial to maximize the use of collision data. Using parameters optimized for each module, the model can predict the afterpulse contributions to the beam-on backgrounds effectively.

%%%%%%%%%%%%%%%%%%%%%%%%%%%%%%%%%%%%%%%%%%%%%%%%%%%%%%%%%%%%%%%%%%%%%%%%%%%%%%%%
\section*{Acknowledgment}
%%%%%%%%%%%%%%%%%%%%%%%%%%%%%%%%%%%%%%%%%%%%%%%%%%%%%%%%%%%%%%%%%%%%%%%%%%%%%%%%
This work was supported by the National Research Foundation of Korea (NRF) grants funded by the Korea government (MSIT) (RS-2021-NR059935 and RS-2025-00560964) and a Korea University grant.

%%%%%%%%%%%%%%%%%%%%%%%%%%%%%%%%%%%%%%%%%%%%%%%%%%%%%%%%%%%%%%%%%%%%%%%%%%%%%%%%
% References
%%%%%%%%%%%%%%%%%%%%%%%%%%%%%%%%%%%%%%%%%%%%%%%%%%%%%%%%%%%%%%%%%%%%%%%%%%%%%%%%
\vspace{0.2cm}
\noindent
\let\doi\relax
\bibliographystyle{ptephy}
\bibliography{main}

%%%%%%%%%%%%%%%%%%%%%%%%%%%%%%%%%%%%%%%%%%%%%%%%%%%%%%%%%%%%%%%%%%%%%%%%%%%%%%%%
%%%%%%%%%%%%%%%%%%%%%%%%%%%%%%%%%%%%%%%%%%%%%%%%%%%%%%%%%%%%%%%%%%%%%%%%%%%%%%%%
\end{document}